\documentclass[a4paper,twocolumn,11pt,accepted=2024-11-26]{quantumarticle}
\pdfoutput=1
\usepackage[utf8]{inputenc}
\usepackage[english]{babel}
\usepackage[T1]{fontenc}
\usepackage{amsmath}
\usepackage{hyperref}
\usepackage[sort&compress,numbers]{natbib}

\usepackage{tikz}
\usepackage{lipsum}

\usepackage{graphicx,mathtools,subcaption}
\usepackage{appendix,xcolor}
\usepackage{lineno}
\usepackage{dsfont}
\usepackage{tikz}
\usetikzlibrary{quantikz}
\newcommand{\ham}{\hat{\mathcal{H}}}

\newcommand{\bm}{\boldmath}

\begin{document}

\title{Efficient preparation of the AKLT State with Measurement-based Imaginary Time Evolution}

\author{Tianqi Chen}
\affiliation{New York University Shanghai; NYU-ECNU Institute of Physics at NYU Shanghai, 567 West Yangsi Road, Pudong New District, Shanghai 200126, China}  
\affiliation{Department of Physics, National University of Singapore, Singapore 117551}
\affiliation{Centre for Quantum Technologies, National University of Singapore, Singapore 117543}
\affiliation{School of Physical and Mathematical Sciences, Nanyang Technological University, Singapore 639798}

\author{Tim Byrnes}
\email{tim.byrnes@nyu.edu}
\affiliation{New York University Shanghai; NYU-ECNU Institute of Physics at NYU Shanghai, 567 West Yangsi Road, Pudong New District, Shanghai 200126, China}  
\affiliation{State Key Laboratory of Precision Spectroscopy, School of Physical and Material Sciences, East China Normal University, Shanghai 200062, China}
\affiliation{Center for Quantum and Topological Systems (CQTS), NYUAD Research Institute, New York University Abu Dhabi, UAE}
\affiliation{Department of Physics, New York University, New York, NY 10003, USA}

\maketitle

\begin{abstract}
  Quantum state preparation plays a crucial role in several areas of quantum information science, in applications such as quantum simulation, quantum metrology and quantum computing. However, typically state preparation requires resources that scale exponentially with the problem size, due to their probabilistic nature or otherwise, making studying such models challenging. In this article, we propose a method to prepare the ground state of the Affleck-Lieb-Kennedy-Tasaki (AKLT) model deterministically using a measurement-based imaginary time evolution (MITE) approach.  By taking advantage of the special properties of the AKLT state, we show that it can be prepared efficiently using the MITE approach. Estimates based on the convergence of a sequence of local projections, as well as direct evolution of the MITE algorithm suggest a constant scaling with respect to the number of AKLT sites, which is an exponential improvement over the naive estimate for convergence.  We show that the procedure is compatible with qubit-based simulators, and show that using a variational quantum algorithm for circuit recompilation, the measurement operator required for MITE can be well approximated by a circuit with a much shallower circuit depth compared with the one obtained using the default Qiskit method. 
\end{abstract}

\section{Introduction}

The last few decades have witnessed tremendous progress in the study of quantum simulation of many-body quantum systems~\cite{Anderson1972,Porras2004,bloch2005ultracold,Brown2006,Bloch2008,Friedenauer2008,bloch2012quantum,Schneider2012,MiyakeKetterle2013,jotzu2014experimental,Schreiber2015,Choi2016,Bordia2017,mazurenko2017cold,salomon2019direct,Cooper2019,Kyprianidis2021observation,shen2023proposal}, one of the earliest proposed practical applications of a quantum computer as suggested by Feynman~\cite{feynman_simulating_1982,lloyd1996universal}.
Recently, with the emergence of noisy intermediate scale quantum (NISQ) era devices~\cite{Preskill2018}, together with improvements in algorithms and hardware, there has been an increasing focus on utilizing such devices to facilitate certain tasks which were beyond the scope of numerical approaches on classical computers.  Examples of models that have been examined are strongly correlated lattice models~\cite{greiner2002quantum,tarruell2018quantum,hensgens2017quantum,byrnes2007quantum}, quantum spin liquids~\cite{Semeghini2021,Giudici2022}, fractional quantum Hall effect~\cite{RahmaniJiang2020}, and models involving Majorana fermions~\cite{huang2021emulating}. The NISQ devices have also been applied in characterizing non-equilibrium phase transition~\cite{Chertkov2023characterizing} as well as dissipative quantum systems involving non-unitary operations~\cite{Schlimgen2021,Schlimgen2022}.

In order to examine the properties of a physical model, the task of state preparation is a crucial step for simulating quantum many-body systems, and has attracted much attention recently~\cite{Ran2020,holmes2020efficient,LinPollmann2021}.  The aim of quantum simulation is to investigate the properties of a particular system, which usually amounts to obtaining the low-lying energy eigenstates of a specified Hamiltonian~\cite{byrnes2021quantum}. Several methods have been proposed for realizing this.  First, in an analogue quantum simulation approach~\cite{Bloch2008,Barredo2016atom,Endres2016atom}, the system is physically cooled down, and the Hamiltonian of interest is realized using experimental techniques such as optical lattices and optical tweezers~\cite{DeutschBrennen2000,Gross2017,Kaufman2021quantum}.  In a digital quantum simulation approach, the quantum phase estimation~\cite{Kitaev1995quantum} algorithm, or other Fourier transform methods~\cite{somma2002simulating}, can be used to read out of the energy eigenstates of the Hamiltonian.  However, in order to study the low energy eigenstates, one must be able to prepare a state with a suitably high overlap with the low energy eigenstates. One approach to prepare this is to use an adiabatic procedure, where one slowly changes the Hamiltonian from a known one to the target Hamiltonian~\cite{Albash2018}.  Another approach that has gained recent popularity are variation approaches, such as the variational quantum eigensolver (VQE)~\cite{Peruzzo2014variational,Tilly2022variational} and the quantum approximate optimization algorithm (QAOA)~\cite{Farhi2014quantum}.  In these approaches, a measurement-feedback approach is used to obtain a low-energy state of a parameterized circuit. Recently, quantum steering has been used to transform an arbitrary initial state to a final target state \cite{RoyGefen2020,HerasymenkoGefen2023}.  Finally, several methods that mimic imaginary time evolution (ITE) have been developed recently.  The variational imaginary time evolution (VITE)~\cite{Mcardle2019variational} method works in a similar way to the measurement-feedback approaches above, and uses a parameterized circuit to find a unitary evolution that best approximates the imaginary time evolution.  Another approach is measurement-based imaginary time evolution (MITE), which uses measurements combined with adaptive unitary operations to result in a stochastic, yet deterministic procedure to prepare eigenstates of Hamiltonians~\cite{MaoByrnes2022,KondappanByrnes2022}.  Other recent approaches which deal with non-unitary time evolution either approximate the target non-unitary operators as unitary ones~\cite{motta2020determining}, or transform them into unitary operators with additional ancilla qubits~\cite{Liu2021three,Chen2024directprobetopologygeometry}.

In all the techniques listed above, finding the ground state of a given Hamiltonian is typically a difficult problem.  For example, Bittel and Kliesch showed that training VQE is NP-hard \cite{bittel2021training}. Methods involving postselection tend to scale badly for large-scale systems due to the low probability of obtaining the desired result \cite{terashima2005nonunitary,liu2021probabilistic,LinPollmann2021}.  Both adiabatic methods and MITE are known to have an exponential convergence time in the typical case, due to an exponentially small gap \cite{Albash2018,MaoByrnes2022}.  
Hence performing state preparation in a large-scale system with a non-trivial wavefunction is an outstanding problem due to the exponential resources required.  This is also true not only for quantum simulation but also for related subfields such as quantum metrology~\cite{giovannetti2011advances,toth2014quantum,you2017multiparameter} and quantum computing~\cite{RaussendorfBriegel2001,Nayak2008,byrnes2012macroscopic}, as these approaches require preparation of a large-scale entangled quantum state.

Recently, there has been some interest in preparing the ground state of the Afflect, Kennedy, Lieb and Tasaki (AKLT) model~\cite{AKLT1987,Affleck1989}.  The AKLT model is a fundamental model in condensed matter physics which has served to understand the physics of 
 fractional excitations at its boundaries~\cite{AKLT1987,Kennedy1990,WhiteHuse1993}, as well as the strongly correlated symmetry protected topological (SPT) phase with a Haldane gap~\cite{Affleck1989}, due to its exactly solvable nature.  Furthermore, the foundational  valence-bond solid (VBS) state~\cite{Knabe1988Energy} serves as the ground state for the spin-1 AKLT model with bilinear-biquadratic interactions. This state holds significance as it presented crucial evidence, supporting Haldane's conjecture~\cite{Haldane1983Nonlinear}, which posits that one-dimensional integer-spin Heisenberg models exhibit a gap. Despite its association with one-dimensional systems, the VBS construction can be extended to host higher-dimensional arbitrary lattices~\cite{AKLT1987,Affleck1988valence}. It is also interesting from the point of view of quantum computing since it was shown to be realizable experimentally with photonics systems using cluster states~\cite{RaussendorfBriegel2001,Briegel2009measurement,WeiAKLT2022}, which is the primary resource of measurement-based quantum computing~\cite{BrowneRudolph2005,BrennenMiyake2008,Kaltenbaek2010optical,Darmawan2012}. The AKLT model also serves as an excellent benchmarking model for testing quantum simulation on NISQ-era devices and further developing associated techniques.  In addition, while the one-dimensional AKLT model is exactly solvable and its ground state can be represented as an MPS, from a technical standpoint, it cannot be generated using a constant-depth local unitary circuit. This is because, in the evolution of such circuits, the correlation between any two local operators becomes negligible beyond the boundaries of a causal light cone~\cite{Smith2022}. As such, several works have recently discussed creating the state on currently available quantum processors~\cite{Murta2022,Smith2022,Chen2023},  including tensor networks-based approaches~\cite{Barratt2021parallel,Slattery2021quantum,Haghshenas2022variational,Miao2023quantum,ZhangChenstack2022,malz2023preparation}, and approaches involving dissipation on a qutrit superconducting array~\cite{wang2023dissipative}. However, those methods either require mid-circuit measurements~\cite{Smith2022}, adiabatic preparation of the state~\cite{Wei2022}, or measurement-based post selection~\cite{Murta2022,Chen2023} or quantum steering~\cite{RoyGefen2020}. As a result, it is also very difficult to extract other important quantities from these approaches such as a direct measurement of observables to compute entanglement entropy without any feed-foward processing~\cite{Smith2022}. Other methods dealing with various non-unitary operators include matrix decompositions~\cite{Mazziotti2022Lindblad} and diagonal operators~\cite{Mazziotti2022Diagonal}. Postselection makes the methods inefficient for longer chains since the probability of a successful preparation reduces drastically with chain length. Hence, compared with the existing preparations, deterministic quantum algorithms with constant depth of quantum circuits that work without the necessity of postselection or complex operations are in demand, such that they are practically implementable for today's NISQ-era quantum processors.

In this work, we propose an efficient, deterministic (i.e. no postselection) method to prepare the ground state of the AKLT model regardless of the system sizes considered. 
Our approach is based upon a unique parallelization of the MITE methods introduced in Refs.~\cite{MaoByrnes2022,KondappanByrnes2022}, where the ground state of a given Hamiltonian can be prepared deterministically, using measurements and adaptive unitary operations. While the AKLT state is well-known to be a simultaneous eigenstate of projection operators, they do not commute with each other, meaning that it cannot be represented as a product states of ground states of the local Hamiltonian. In addition to the fundamental interest of the AKLT state, the model is interesting to consider in the context of the MITE algorithm due to the very high preparation efficiency that can be attained.  For a general Hamiltonian the ground state preparation for MITE requires an exponential time overhead with system size \cite{MaoByrnes2022}. Remarkably, in the case of AKLT ground state, this can be reduced to constant time with respect to the chain length $ N $.  This exponential speedup is thanks to the specific properties of the AKLT model that makes this possible, despite the large Hilbert space and the complexity of the quantum many-body wavefunction.  We also show that our approach can be implemented not only on qudit-based devices directly, but also on qubit-based devices.  The key procedure of MITE, which involves a measurement in the energy eigenbasis of a given Hamiltonian, can be implemented efficiently using a variational optimization algorithm for circuit recompilation.

\section{The AKLT model}
\label{sec:modelsandmethods}

\begin{figure}[t]
	\centering	\includegraphics[width=1.0\columnwidth,draft=false]{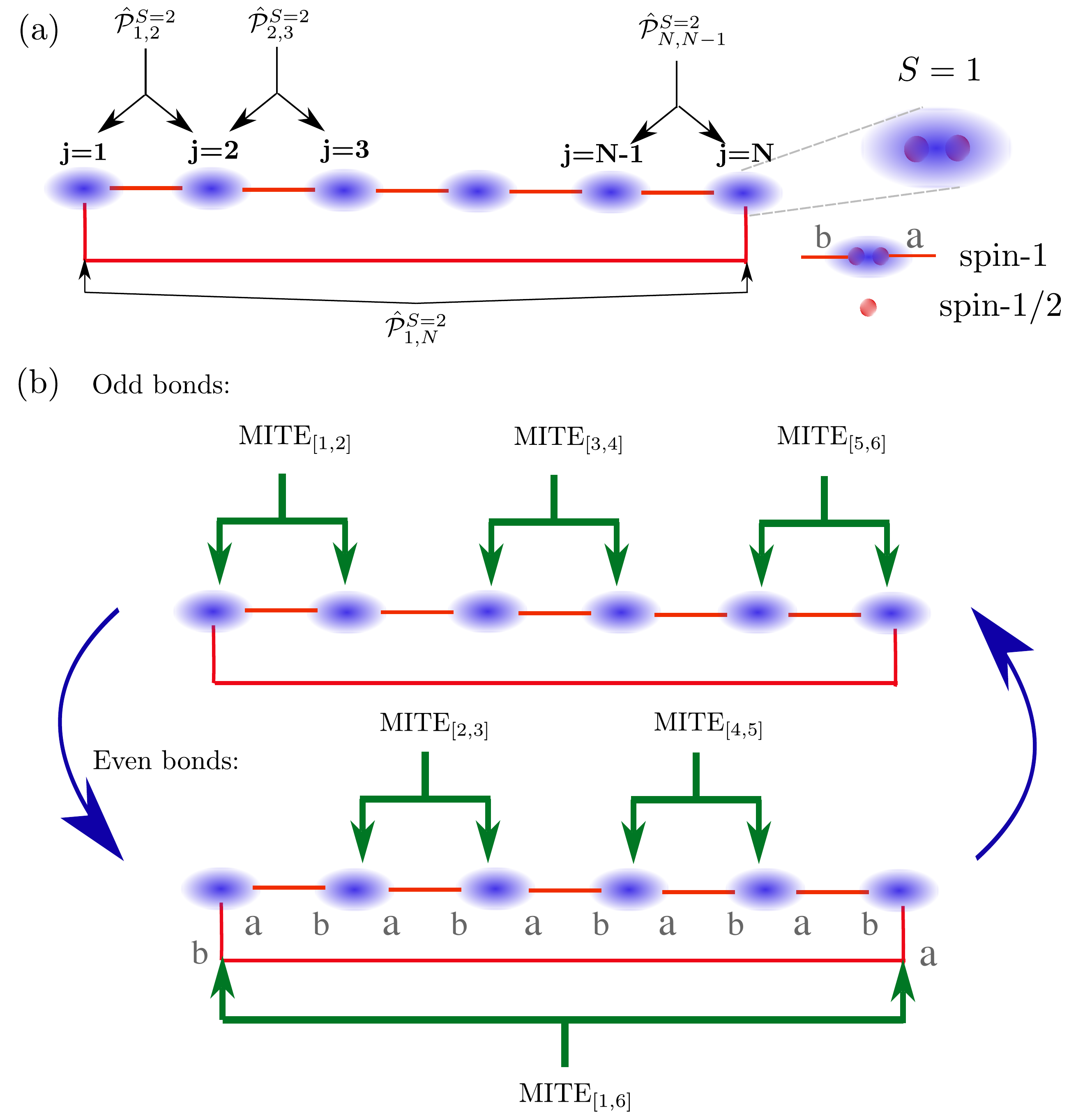}
	\caption{(a) The $ N $-site AKLT model. Each spin-$1$ site is represented by an oval, which may be either a spin-$1$ particle or two coupled spin-$1/2$ particles.  The lines between the sites represent the interactions, indicating the PBC geometry. 
 The AKLT ground state is the common zero eigenstate [Eq.~\eqref{eq:AKLTHamiltonianPBC}] of the $ S=2$ projectors $\hat{\mathcal{P}}_{j,j+1}^{S=2} $ on adjacent sites, indicated by the branched arrows.  (b) State preparation of the AKLT state using MITE for $N=6$. The MITE subroutine, denoted by  $\text{MITE}_{[m,n]}$ (forked arrows) is executed simultaneously on neighboring pairs of sites, firstly for the odd bonds (upper panel) and subsequently for the even bonds (lower panel). The procedure repeats until convergence.  }
	\label{fig:AKLTstate}
\end{figure}

We consider a one-dimensional AKLT model, which consists of a linear chain of $N$ spin-$1$ particles. The Hamiltonian for such an AKLT model is
\begin{align}
  \label{eq:AKLTHamiltonianPBC}
\ham  &=\sum_{j=1}^{N} \hat{\mathcal{P}}_{j,j+1}^{S=2}
\end{align}
where periodic boundary conditions (PBC) is used throughout this work, and 
\begin{align}
  \label{eq:spin2projector}
    &\hat{\mathcal{P}}_{j,j+1}^{S=2}=\frac{1}{24}\left(\mathbf{S}_j+\mathbf{S}_{j+1}\right)^2 \left[\left(\mathbf{S}_j+\mathbf{S}_{j+1}\right)^2-2\hat{I}\right] \\ \nonumber &=\frac{1}{2}\left[\mathbf{S}_j \cdot \mathbf{S}_{j+1}+\frac{1}{3}\left(\mathbf{S}_j \cdot \mathbf{S}_{j+1}\right)^2+\frac{2}{3} \hat{{I}}\right],
\end{align}
is the projector on the $ S = 2 $ subspace, where we used 
$ \mathbf{S}_j \cdot \mathbf{S}_{j}=2 \hat{I}$.  Here we defined $\mathbf{S}_{j}=(S_j^x,S_j^y,S_j^z)$  for $j=[1, N]$, where
\begin{align}
  \label{eq:spin1operator}
  &S_j^x=\frac{1}{\sqrt{2}}
  \begin{pmatrix}
    0&1&0 \\ 
    1&0&1 \\ 
    0&1&0 
  \end{pmatrix}, 
  S_j^y=\frac{i}{\sqrt{2}}
  \begin{pmatrix}
    0&-1&0 \\ 
    1&0&-1 \\ 
    0&1&0 
  \end{pmatrix}, \nonumber \\
  &  S_j^z=
  \begin{pmatrix}
    1&0&0 \\ 
    0&0&0 \\ 
    0&0&-1 
  \end{pmatrix}.
\end{align}
Throughout this work, we consider periodic boundary conditions (PBC), such that the site $j=N+1$ is equivalent to site $j=1$. 

The ground state of the one-dimensional AKLT model with PBC has no degeneracy \footnote{For the one-dimensional AKLT model with open boundary conditions, it has a four-fold degeneracy because of the two separate spin-$1/2$'s located at each boundary.}. The ground state has an eigenvalue equal to zero exactly
\begin{align}
\ham | \text{AKLT}  \rangle = 0 .
\end{align}
We shall henceforth call this non-degenerate ground state of the AKLT model \eqref{eq:AKLTHamiltonianPBC} the ``AKLT state''.  The ground state also satisfies for all $ j $~\cite{WeiAKLT2022}
\begin{align}
 \hat{\mathcal{P}}_{j,j+1}^{S=2} | \text{AKLT}  \rangle = 0 ,
 \label{commonproj}
\end{align}
which will be important in our case. This implies that for each neighboring sites, the total net spin should consist of $S=0$ and $S=1$ states.  Note that the projectors on adjacent pairs of sites do not commute, $ [ \hat{\mathcal{P}}_{j-1,j}^{S=2},\hat{\mathcal{P}}_{j,j+1}^{S=2} ]  \ne 0 $, even though the ground state is a common eigenstate of all the projection operators. Here, we remark that even though the adjacent pairs of the terms do not commute with each other, it is not required to perform a Suzuki-Trotter decomposition~\cite{Suzuki1990fractal} of Eq. (\ref{eq:AKLTHamiltonianPBC}) as done in our previous work~\cite{MaoByrnes2022}, which suggests that state preparation may be performed more efficiently. We note that since an AKLT state can be written exactly as a matrix product state (MPS)~\cite{schollwock2011density}, this also implies that for each pair of sites, the local reduced density matrix has a minimum eigenvalue of $0$.

Instead of using genuine $ S = 1 $ particles, one may also create the AKLT state by using two $ S=1/2 $ particle per site, as shown in Fig.~\ref{fig:AKLTstate}. While we will not directly use this in our procedure, we summarize some of these properties in this formulation.  The AKLT state can be prepared by first preparing singlet states between the sites
\begin{align}
    \bigotimes_{j=1}^N \left( |0 \rangle_{j,a} |1 \rangle_{j+1,b} -|1 \rangle_{j,a} |0 \rangle_{j+1,b}  \right) ,  
    \label{singlets}
\end{align}
where $ l \in \{ a,b \} $ label the two subspins on each site. Then applying a projection operator for all $ j \in [1,N] $
\begin{align}
\hat{\mathcal{P}}_{j}^{S=1} = \frac{4}{3} \bm{\sigma}_{j,a} \cdot \bm{\sigma}_{j,b} 
\label{spinoneproj}
\end{align}
where 
\begin{align}
\bm{\sigma}_{j,l} =(\sigma_{j,l}^x,\sigma_{j,l}^y,\sigma_{j,l}^z) 
\label{spinhalf}
\end{align} 
for $j=[1, N] $ and
\begin{align}
  \label{eq:spin1operator}
  &\sigma_{j,l}^x=\frac{1}{2}
  \begin{pmatrix}
    0&1 \\ 
    1&0
  \end{pmatrix},
 \sigma_{j,l}^y =\frac{i}{2}
  \begin{pmatrix}
    0&-1 \\ 
    1&0
  \end{pmatrix},
 \sigma_{j,l}^z =
  \begin{pmatrix}
    1&0 \\ 
    0&-1 
  \end{pmatrix},
\end{align}
one achieves the ground state of the Hamiltonian from Eq.~\eqref{eq:AKLTHamiltonianPBC}.  In Eq.~\eqref{singlets}, the states are defined as the eigenstates of the Pauli-$z$ operator $ \sigma_{j,l}^z | n \rangle_{j,l} = (-1)^n | n \rangle_{j,l}$ for $n \in \{0, 1 \}$.
The projective (i.e. measurement) nature of Eq.~\eqref{spinoneproj} suggests that it might be compatible with MITE, but in fact the random nature of quantum measurements, and the specific initial state from Eq.~\eqref{singlets} that is required makes a reliable preparation more difficult.   In the following section, we explain an approach that is more robust and efficient.

\section{The MITE algorithm}
\label{sec:mite}

\subsection{Overall scheme}

We now describe a deterministic procedure to generate the AKLT state using measurements and unitary operations.  To this end, we use the measurement-based imaginary time evolution (MITE) approach~\cite{MaoByrnes2022}, adapted towards the specific state of interest in this case. The MITE algorithm uses a combination of quantum measurements with adaptive unitary transformations to deterministically prepare an eigenstate of a given Hamiltonian.  In the case of the AKLT state, since it is the common eigenvalue of the $ \hat{\mathcal{P}}_{j,j+1}^{S=2} $ operators as shown in Eq.~\eqref{commonproj}, this allows for a more efficient procedure than simply applying the MITE algorithm to the Hamiltonian from Eq.~\eqref{eq:AKLTHamiltonianPBC}.  

In imaginary time evolution, the goal is to amplify the ground state, such that we obtain the AKLT state from an arbitrary initial state $|\psi_0\rangle$
\begin{align}
  \label{eq:ITE}
  &\lim_{\tau\rightarrow\infty}e^{-\ham \tau}|\psi_0\rangle=|\text{AKLT} \rangle .
\end{align}
Due to the fact that the Hamiltonian is the sum of all local projectors on the $S=2$ subspace, the property of Eq.~\eqref{commonproj} can therefore be applied to the Hamiltonian, and we observe that we also have
\begin{align}
\lim_{\tau\rightarrow\infty} \prod_{j=1}^N e^{-\tau \hat{\mathcal{P}}_{j,j+1}^{S=2} } |\psi_0\rangle=|\text{AKLT} \rangle .
\end{align}
Note that this is a specific relation for the AKLT state since $ \hat{\mathcal{P}}_{j,j+1}^{S=2} $ do not all commute. Our approach to MITE will therefore be to perform a MITE subroutine on each pair of sites $j$ and $j+1$ locally, and sweep over all pairs of sites for many iterations until the fidelity is converged to one with the target state. The purpose of the sweeps is to converge towards a consistent state such as to satisfy Eq.~\eqref{commonproj} for all $ j $. We illustrate the sequence for the preparation of the AKLT state in Fig.~\ref{fig:AKLTstate}(b). The MITE subroutine is simultaneously applied to each pair of sites with the same $ j $ parity.  For example, for the $ N = 6$ case the $ (j, j+1) = (1,2), (3,4), (5,6) $ are applied at the same time.  This is followed by $ (j, j+1) = (2,3), (4,5), (6,1) $.  This parallelization is possible because these operations all commute, as they do not share common spins.

\subsection{MITE subroutine on a pair of sites}

We now describe how the MITE subroutine works on a pair of sites $(j, j+1) $.  One of the key ingredients of MITE is to perform a measurement in the eigenbasis of the desired Hamiltonian~\cite{MaoByrnes2022}.  In the standard MITE algorithm, this would be the eigenstates of the Hamiltonian from Eq.~\eqref{eq:AKLTHamiltonianPBC}.  For a system with a large Hilbert space, this may require long convergence times, since in general the convergence scales with the Hilbert space dimension~\cite{MaoByrnes2022}. Here, we propose a more efficient method, taking advantage of the property as given in Eq.~\eqref{commonproj}.  

We instead form the measurement operators in the eigenbasis of $ \hat{\mathcal{P}}_{j,j+1}^{S=2} $, which are
\begin{widetext}
\begin{align}
  \label{eq:weakmeasurementoperator}
 &M_{j,j+1}^q=  \frac{1}{2} \sum_{S=0}^2 \sum_{m=-S}^{m=S}    \left(\cos\epsilon E_{S,m} - (-1)^q \sin \epsilon E_{S,m} \right) |S, m  \rangle_{j,j+1} \langle S, m |_{j,j+1} 
\end{align}
\end{widetext}
where $q \in \{0,1 \} $ labels the measurement outcomes. In Ref.~\cite{MaoByrnes2022} it is described how to realize a measurement of this form using an ancilla qubit prepared in the state $ | + \rangle_a = (|0 \rangle_a + |1 \rangle_a )/\sqrt{2} $.  Applying a unitary evolution $ \exp (-i\epsilon\hat{\mathcal{P}}^{S=2}_{j,j+1} \otimes \sigma_a^{y} ) $ and measuring the ancilla qubit in the $ z $-basis produces the measurement in Eq.~\eqref{eq:weakmeasurementoperator}.   Here, $ |S, m \rangle_{j,j+1} $ 
are angular momentum eigenstates with total spin $ S $ and projection $ m $ and are eigenstates of $ \hat{\mathcal{P}}^{S=2}_{j,j+1} $.  The eigenstates consist of two coupled $ S=1 $ particles at sites $ j $ and $ j+1$.  The eigenvalues of the projection operator are given by 
\begin{align}
E_{S,m} = \delta_{S,2} ,
\label{energyeigen}
\end{align}
i.e. it is 1 if $ S= 2 $ and 0 otherwise.

In a single MITE subroutine, a sequence of measurements given by Eq.~\eqref{eq:weakmeasurementoperator} are performed, where the outcomes $ q \in \{0,1 \}$ occur randomly according to Born probabilities. For a particular sequence, if there are a total of 
$k_0$ counts of $q=0$ and $k_1$ counts of $q=1$, the resultant state is
\begin{align}
  \label{eq:measurementsequences}
  & (M_{j,j+1}^0 )^{k_0} (M_{j,j+1}^1 )^{k_1}
  |\psi_0 \rangle \\ \nonumber &=\sum_{S,m}  A_{k_0 k_1} \left(\epsilon E_{S,m} \right) \langle E_{S,m} | \psi_0 \rangle | E_{S,m} \rangle
\end{align}
where 
\begin{align}
A_{k_0 k_1} (\chi)  = \cos^{k_0} (\chi + \pi/4) \sin^{k_0} (\chi - \pi/4) 
\end{align}
is an amplitude function which has a Gaussian form for a large number of measurements.  The sequence of measurements therefore causes a collapse of the state in energy space, peaked at the value 
\begin{align}
  \label{eq:peakofA}
  &  E_{k_0 k_1}^{\rm peak}=\frac{1}{2\epsilon }\arcsin{\frac{k_1-k_0}{k_0+k_1}} .
\end{align}
Since in our case, the only energy eigenstates are $ E = 0, 1 $ according to Eq.~\eqref{energyeigen}, there will be an amplitude gain of the desired $ E=0 $ states as long as $ E_{k_0 k_1}^{\rm peak} < 1/2 $, the midway point between the energy levels.

Applying the measurement sequence [Eq.~\eqref{eq:measurementsequences}] randomly converges to either the target $ E = 0 $ states (with $ S = 0, 1 $), or the $ E = 1 $ states (with $ S = 2$).  In order to make a deterministic procedure to produce the AKLT state, we require an 
adaptive operation, where we introduce a conditional unitary operator applied to the measurement outcomes. By monitoring the location of the Gaussian peak [Eq.~\eqref{eq:peakofA}], if its value is higher than a pre-chosen energy threshold $E_{\rm th}$ where $E_{\text{th}}=\epsilon/\eta $, $\epsilon$ is the interaction time, and $\eta$ is an adjustable modification factor, then we apply a corrective unitary to the state. If $E^{\rm peak}_{k_0 k_1} $ falls into the desired range $E^{\rm peak}_{k_0 k_1} < E_{\rm th} $, then no further operation is executed.  To make this concrete, we define the adaptive unitary operation as
\begin{align}
  \label{eq:remedyunitary}
  &U_{j,j+1}=\begin{cases}
    \hat{I} & \quad \text{if}\,\,E^{\rm peak}_{k_0 k_1} < E_{\rm th} \\ 
    U^C_{j,j+1} & \quad \text{otherwise} ,  
  \end{cases}
\end{align}
where $ \hat{I}$ is the identity operator. Combining the adaptive unitary with the measurement, we hence perform the sequence after the $T$th measurement as
\begin{align}
  \label{eq:evolutionflow}
  &|\psi_T \rangle \rightarrow |\psi_{T+1}\rangle = \frac{U_{j,j+1} M_{j,j+1}^{q} |\psi_T \rangle }{\sqrt{\langle \psi_T | M_{j,j+1}^{q,\dagger} M_{j,j+1}^{q}| \psi_T \rangle}}
\end{align}
where the choice of which unitary that is made in Eq.~\eqref{eq:remedyunitary} is made by keeping track of the counts $ k_0 $ and $ k_1$.  Once the operator $ U^C $ is applied, the counts are reset to zero $ k_0 = k_1 = 0 $.  

The particular choice of adaptive unitary that is made has a large amount of freedom, and can be any operator that creates a transition between the $ E = 0 $ and $ E = 1$ energy states.  We choose a form that rotates the spins randomly on each of the sites
\begin{align}
  \label{eq:correctionunitary}
  U_{j,j+1}^C = & \exp\left[2\pi i \left(a_{j}^{x}S_{j}^x + a_{j}^{y}S_{j}^y + a_{j}^{z}S_{j}^z \right)\right] \nonumber \\
  & \otimes \exp\left[2\pi i \left(a_{j+1}^{x}S_{j+1}^x + a_{j+1}^{y}S_{k}^y + a_{j+1}^{z}S_{k}^z \right)\right] 
\end{align}
where the $a_{j}^{\alpha} \in [0,1] $ for $ \alpha\in \{ x,y,z \}$ a randomly generated real number.  

For each pair of sites $(j,j+1)$, the procedure as described in Eq.~\eqref{eq:evolutionflow} is repeated up to $N_{\text{iter}}$ times, or until convergence on the MITE subroutine is attained. Convergence is attained in a sequence of measurements when $ E^{\rm peak}_{k_0 k_1} < E_{\rm th} $ for many iterations. 
The MITE subroutine on separate spins can be straightforwardly parallelized, since the updates from Eq.~\eqref{eq:evolutionflow} can be performed completely locally.  For example, in the $ N = 6 $ case shown in Fig.~\ref{fig:AKLTstate}(b), the MITE subroutine is applied to odd and even $ j $ for the pairs of sites involved in the measurement shown in  Eq.~\eqref{eq:weakmeasurementoperator}.  Iterating the MITE subroutines on the odd and even $ j $ pairs of sites, the state converges to the AKLT state. 

\section{Numerical simulations}

\subsection{The AKLT state preparation with spin-$1$ particles}
\label{sec:foursite}

In this section, we show numerical results of the performance for the AKLT state preparation with spin-$1$ particles.  This is a direct preparation of the AKLT state in the spin-$1$ representation of Fig.~\ref{fig:AKLTstate}. This can be physically implemented using qutrits \cite{Ruben2018,WangKais2020} on various platforms such as trapped ions~\cite{Hrmo2023native}, superconducting quantum processors~\cite{Martinis2009}, silicon carbide~\cite{soltamov2019excitation}, as well as silicon-photonic integrated circuits~\cite{chi2022programmable}. 

The initial state for the evolution is chosen to be:
\begin{align}
  \label{eq:spin1initialstate}
  &|\psi_0\rangle=\bigotimes_{j=1}^{N}  |S=1, m=1\rangle_{j} 
\end{align}
where $ |S=1, m\rangle $ are the eigenstates of $S^z$.  We choose this state as an easily preparable initial state experimentally, although the MITE routine is generally not sensitive on the initial conditions.  

To measure the performance, we use the state fidelity with respect to the AKLT state 
\begin{align}
F_{\rm tot}=|\langle \text{AKLT} | \psi_T \rangle|^2. 
\end{align}
We also compute the partial fidelities for each pair of sites $(j,j+1)$, defined as 
\begin{align}
  \label{eq:partialfidelity}
    F_{j,j+1} = \langle \psi_T | (\hat{I}- \hat{\mathcal{P}}^{S=2}_{j,j+1} ) | \psi_T \rangle .  
\end{align}
This calculates the fidelity of successfully converging to the zero eigenvalue state for a single projector $ \hat{\mathcal{P}}^{S=2}_{j,j+1} $.  Here,  $|\psi_T\rangle$ is the state with the update rule as shown in Eq.~\eqref{eq:evolutionflow} 
during the MITE subroutine, after $ T $ measurements are made.   

\begin{figure}[t]
    \centering
\includegraphics[width=1.0\columnwidth,draft=false]{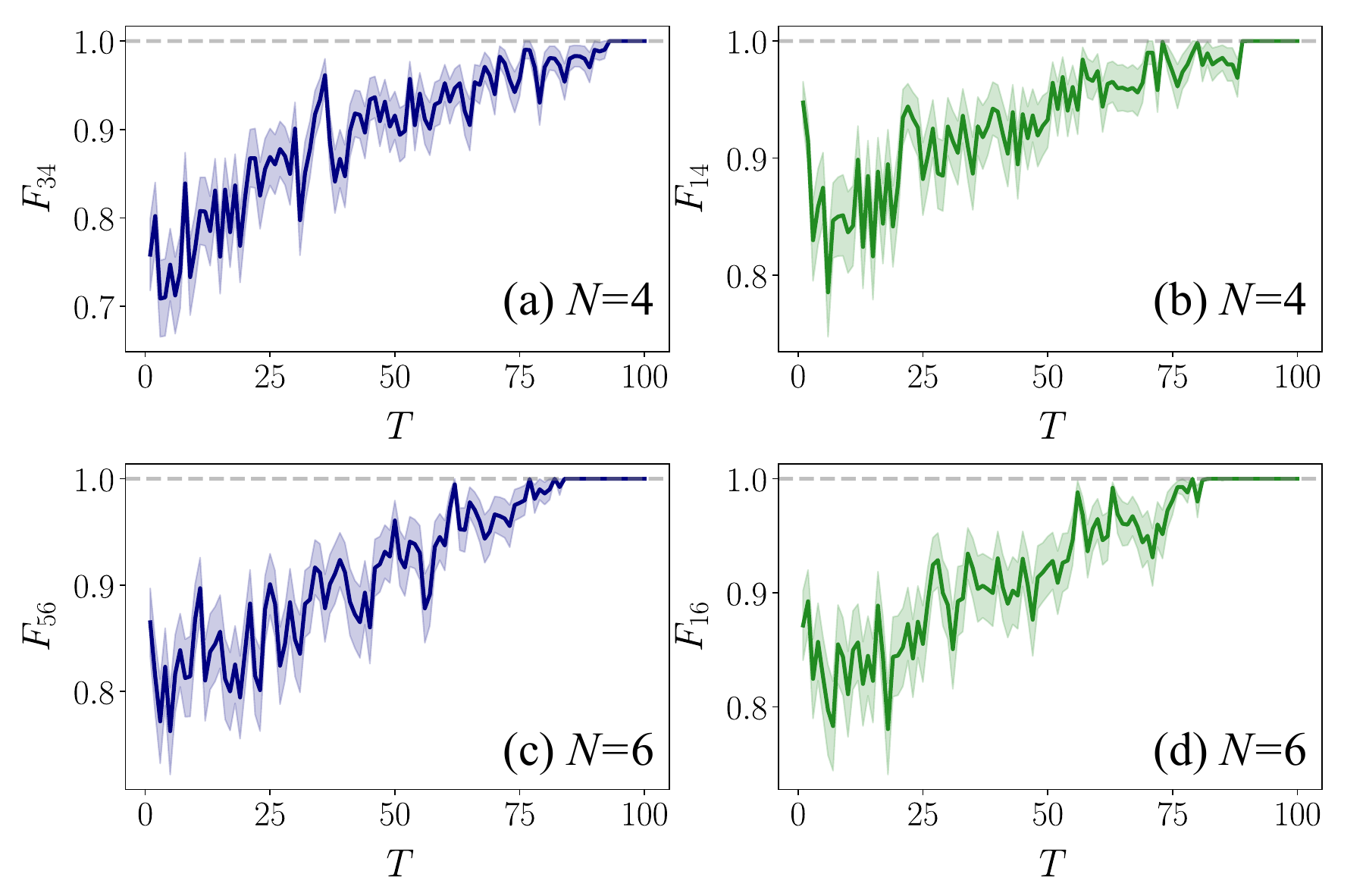}
    \caption{Partial fidelities using the spin-$1$ representation {between odd pairs of sites} (a) $N=4$ ($F_{\rm 3,4}$ as an example), (c) $N=6$ ($F_{\rm 5,6}$ as an example) and between even pairs of sites (b) $N=4$ ($F_{\rm 1,4}$ as an example), (d) $N=6$ ($F_{\rm 1,6}$ as an example).  The darker curves are the averaged partial fidelities calculated over $100$ runs. The shaded area indicates the standard deviations. {The data points are} recorded for each measurement step $T$ within each MITE subroutine up to $T=100$. For all panels, the interaction time is chosen to be $\epsilon=0.5$. The factor $\eta$ is chosen to be $\eta=4$ such that $E_{\text{th}} = 1/8$.  }
    \label{fig:partialfidelity}
\end{figure}

The partial fidelities are shown in Fig.~\ref{fig:partialfidelity}. 
Averaging over 100 runs, we observe that there is a convergence of the partial fidelities to unity typically within $T < 100$ measurements, for all chain lengths that we have tried. Both odd and even parity site pairs converge with a similar number of MITE steps. The trajectory for each particular run is different, due to the stochastic nature of quantum measurements, but every run converges to a state with $ F_{j, j+1}=1$.  We expect that the convergence occurs at similar timescales regardless of the chain length $N $ because the dimension of the MITE subroutine for a pair is constant in each case.  Specifically, there are two sites involved with a dimension of $ 3^2 = 9$.  Due to the key property of Eq.~\eqref{commonproj}, the convergence of each MITE subroutine is independent of the chain length.  

\begin{figure}[t]
	\centering
	\includegraphics[width=1.0\columnwidth,draft=false]{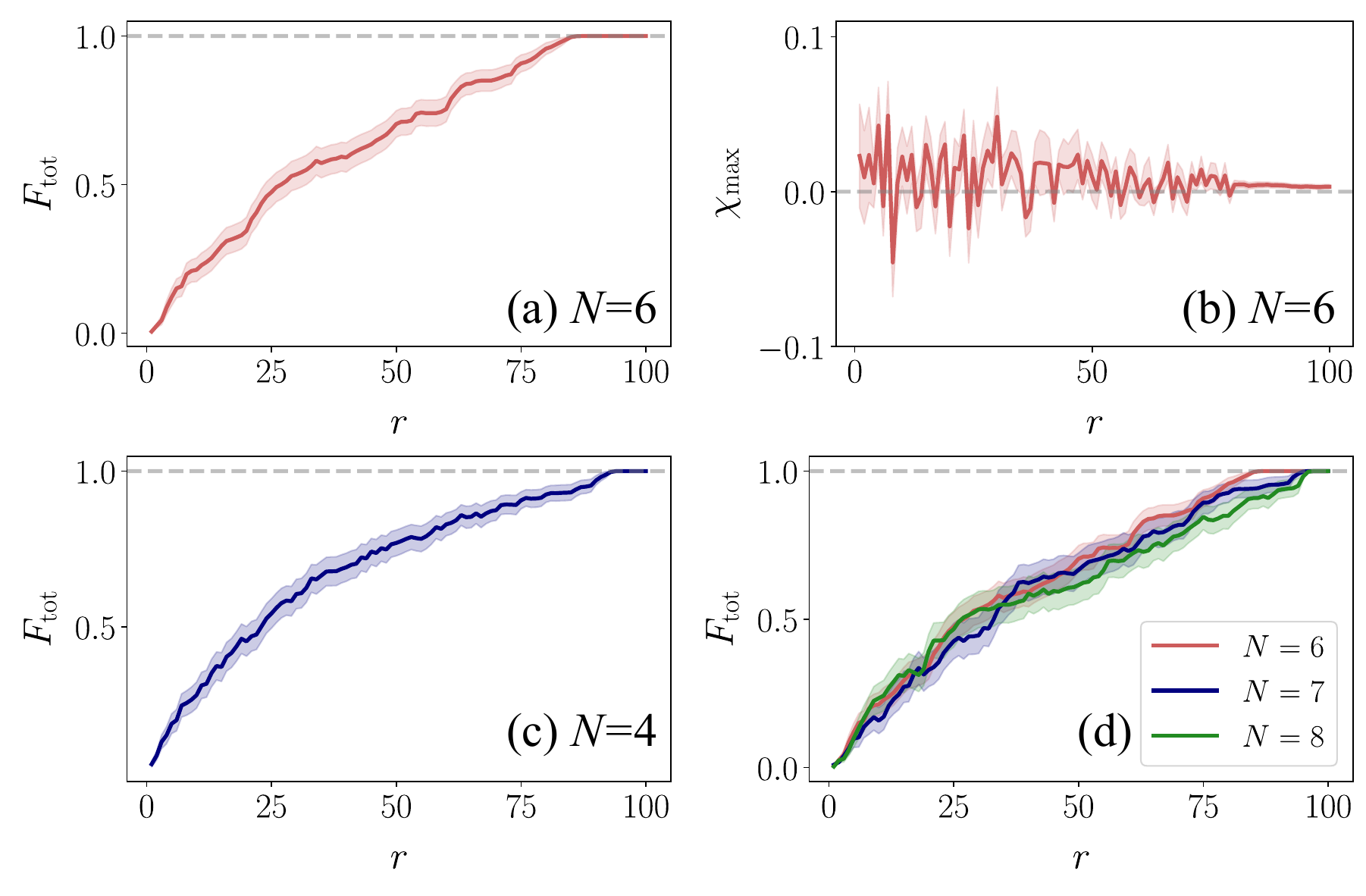}
	\caption{AKLT state preparation using the spin-$1$ representation. The total state fidelity $F_{\rm tot}$ as a function of MITE subroutine round $r$ with (a) $N=6$, and (c) $N=4$. For each $r$, MITE routines are performed throughout the whole system. (b) The peak value of the amplitude function as a function of time with $N=6$. The data points are recorded within the iteration for each MITE subroutine round $r$ throughout the whole system up to $r=100$. (d) Comparison of the total state fidelity $F_{\rm tot}$ for $N=6$ (red), $7$ (blue) and $8$ (green). For all panels, we have used $N_{\rm iter}=30$, and the interaction time is chosen to be $\epsilon=0.5$. The darker curves are the averaged {quantities} calculated over 100 runs. The shaded area indicates the standard deviations. The factor $\eta$ is chosen to be $\eta=4$ such that $E_{\text{th}} = 1/8$.
    }
	\label{fig:N4statepreparation}
\end{figure}

The fidelities with respect to the AKLT state are shown in 
Fig.~\ref{fig:N4statepreparation}(a)(c)(d) for various chain lengths.  
Here we plot the total fidelity after $ r $ complete rounds of applications of MITE subroutines applied to the whole chain, as shown in Fig.~\ref{fig:AKLTstate}(b). 
Similarly to the partial fidelities in Fig.~\ref{fig:partialfidelity}, we see that the AKLT state is prepared within $r = 100$ rounds, based on the average total fidelity calculated over $100$ runs.
Remarkably, there is little dependence of the number of rounds on different chain lengths [Fig.~\ref{fig:N4statepreparation}(d)], despite the exponential difference in Hilbert space dimensions.    

We interpret this in the following way.  Consider a MITE subroutine acting on sites $ j$ and $ j +1 $.  While the MITE evolution deterministically evolves the state towards the eigenstate given in Eq.~\eqref{commonproj}, it may disrupt the eigenvalue relation of its nearby operators, such as $  \hat{\mathcal{P}}_{j-1,j}^{S=2} , \hat{\mathcal{P}}_{j+1,j+2}^{S=2} $.  The disruption is however limited to the vicinity of the sites $ j, j+1 $ due to the short-ranged correlations of the AKLT state.  Due to the limited range of the disruption due to the MITE operations, the chain length is not the relevant factor determining the convergence.  Different parts of the chain effectively act independently, and convergence is attained once the state independently finds the state that is a common zero eigenvalue of all the projectors in Eq.~\eqref{commonproj}.  

\begin{figure}[t]
	\centering
	\includegraphics[width=1.0\columnwidth,draft=false]{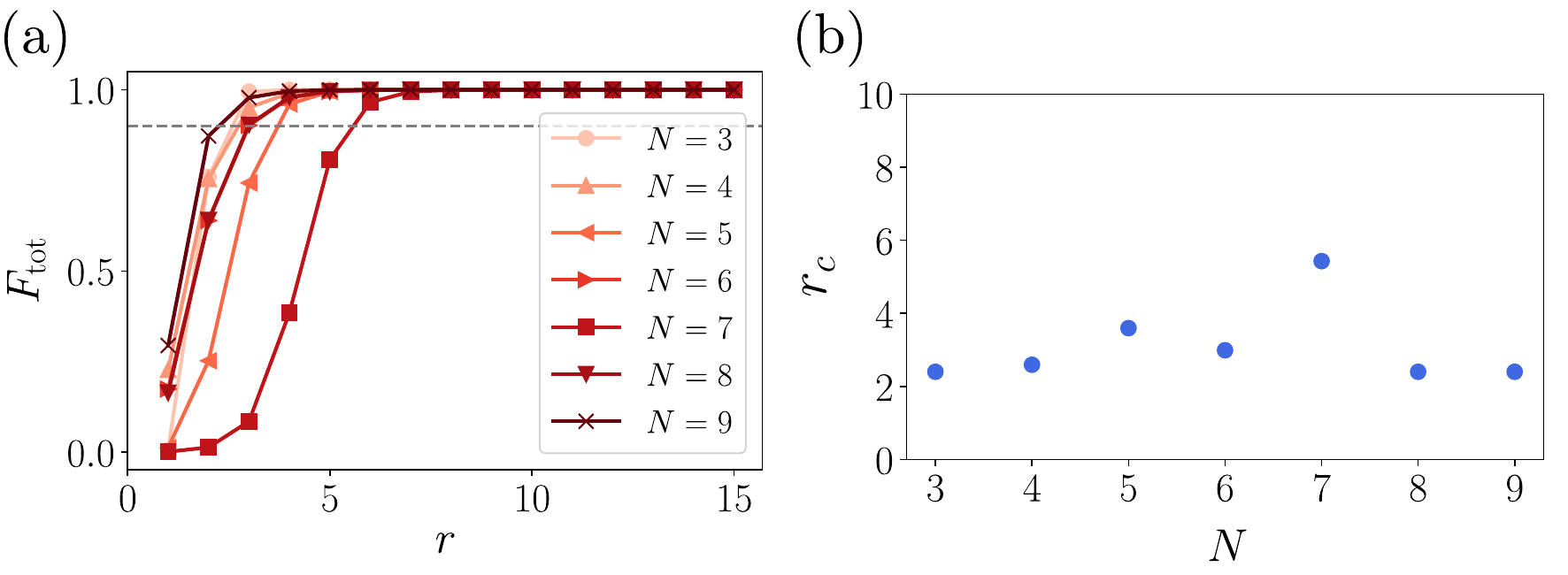}
	\caption{Target AKLT state convergence using direct applications from Eq.~\eqref{eq:directapplication}: (a) the total fidelity $F_{\rm tot}$ as a function of direct application rounds $r$ up to $r=15$. The grey dashed line represents $F_{\rm tot}=0.9$. Lighter to darker colors indicate different system size from $N=3$ to $N=9$. (b) the critical value of rounds $r_c$ when $F_{\rm tot}=0.9$ as a function of system size $N$. All critical values of rounds $r_c$ were obtained via interpolation of the data from panel (a) by numerically solving $F_{\rm tot}(r)=0.9$.  }
	\label{fig:convergence}
\end{figure}

To test this hypothesis, we plot the convergence of the fidelity of the following state
\begin{align}
\label{eq:directapplication}
   \left[  \left( \prod_{j\in \text{even} }  (\hat{I}- \hat{\mathcal{P}}_{j,j+1}^{S=2}) \right) \left( \prod_{j\in \text{odd} }  (\hat{I} - \hat{\mathcal{P}}_{j,j+1}^{S=2})  \right)  \right]^r | \psi_0 \rangle 
\end{align}
with the AKLT state.  Here, $ | \psi_0 \rangle $ is 
the state as in Eq.~\eqref{eq:spin1initialstate} but the eigenstates of $ S^x$.   This may be considered to by a simplified version of our MITE-based procedure, where we remove the stochastic aspect of the MITE subroutines and focus on the convergence of the common projections. The smaller numerical overhead allows us to go to larger $N $.   As can be seen in Fig.~\ref{fig:convergence}(a),  for different system sizes, the state converges to the target AKLT state within $8$ rounds.  In Fig.~\ref{fig:convergence}(b) we plot the number of rounds required to attain a fidelity of 0.9.  We see that the number of rounds remains approximately constant for all chain lengths $ N $.  The particular case $ N = 7 $ shows a higher number of rounds due to the slow increase in fidelity in the early rounds, as see in Fig.~\ref{fig:convergence}(a).  For this case, the initial state appears to be a particularly poor choice, requiring a longer time to convergence.  The convergence confirms our hypothesis and suggests that in our procedure, the total number of rounds $r$ for the convergence of the target state should remain constant with the system size $N$.

In Fig.~\ref{fig:N4statepreparation}(b), we also plot the average peak value computed using Eq.~\eqref{eq:peakofA} which is recorded at the end each round. The average peak values are calculated over $100$ runs. The dashed horizontal lines correspond to the zero energy eigenvalues.  We see that the time evolutions eventually almost converge to the target zero energy values around $r \approx 80$ (the final peak value is of the magnitude $10^{-3}$), after an initially chaotic regime. For an interaction time  $\epsilon=0.5$ the sequence shows good convergence but for smaller values we find the convergence is slower.  This is as expected since a shorter interaction time corresponds to a weaker measurement, and the time taken to collapse takes longer. In short, our MITE approach proves to be an efficient and deterministic quantum algorithm to prepare the AKLT state, where the number of rounds required for the total procedure remains constant even for a moderately large system size.

\begin{figure}[t]
	\centering
	\includegraphics[width=1.0\columnwidth,draft=false]{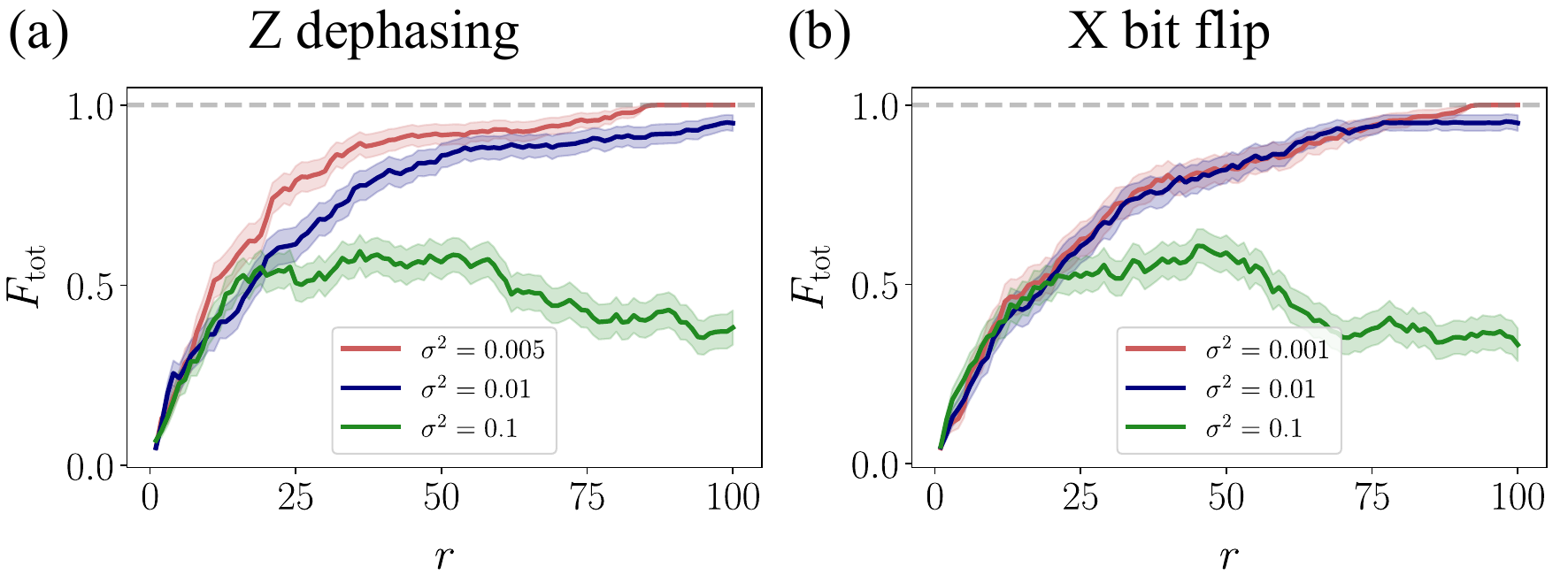}
	\caption{Effect of noise on the AKLT state preparation. (a) $Z$ dephasing noise by adding a rotation $Z$ gate on each site: $\prod_{j=1}^N \exp{[i \xi_j S_j^z]}$, and (b) bit flip noise by adding a rotation $X$ gate on each site: $\prod_{j=1}^N \exp{[i \xi_j S_j^x]}$. Here, we set $N=4$, and for each site, $\xi_j$ is randomly drawn from a normal distribution $\exp{[-\xi^2/\sigma^2]}$. The noise is added at the beginning of each MITE subroutine round $r$, and the data is averaged over $100$ runs.}
	\label{fig:dephasing}
\end{figure}

Finally, in Fig.~\ref{fig:dephasing}, we illustrate the effect of noise on the AKLT state preparation with the MITE procedure by showcasing two different types of noise: 
$Z$ dephasing and the $X$ bit flip noise. 
The noise is modelled by adding a random rotation gate $\prod_{j=1}^N \exp{[i \xi_j S_j^{\nu}]}$ at each MITE subroutine round $r$, where $\xi_j$ is a random number drawn from a normal distribution $\exp{[-\xi^2/\sigma^2]}$ for each site $j$, and $\nu=z (x)$ for dephasing (bit flip) noise. Such a random phase rotation is equivalent to Lindblad dephasing when averaged over many trajectories \cite{pyrkov2014full}.  
We see that for both dephasing and bit flip noise, when $\xi_j$ is significantly small ($\sigma^2$ is close to zero), the MITE method is still robust and is able to converge to a successful preparation of the target AKLT state within $r<100$ [Fig.~\ref{fig:dephasing}]. As for the intermediate strength of the order $\sigma^2 \approx 10^{-2}$, the total fidelity can still reach almost $90 \%$. For all strengths of $\sigma^2$ considered, the $X$ bit flip noise more strongly suppresses $F_{\rm tot}$ during the early stage of the algorithm [Fig.~\ref{fig:dephasing}(b)] compared with the $Z$ dephasing effect.  The MITE approach tends to work robustly under such noise since it is a measurement-based method.  Once it has converged to the correct state, the states tend to be ``locked'' at this value with fidelities equal to 1.  The stochastic iterative approach of MITE means that the particular trajectory that the initial state takes towards the final state is largely irrelevant ---  there are many paths that MITE can take towards obtaining the final result.  Therefore, the presence of small errors along the way do not affect the overall convergence.  However, the noise must not be too large such as to remove the convergence altogether.  In the case of our simulations this threshold appears to take place at noise levels of $ \sigma^2 \sim 10^{-2} $.

\subsection{The AKLT state preparation with qubits}
\label{sec:threesite}

While the standard form of the AKLT Hamiltonian is written with respect to spin-1 particles, typical quantum computers work with qubits, which may be considered to be spin-$1/2$ particles. In this section, we demonstrate that our methods are compatible with qubit-based quantum devices, such as transmon qubit superconducting quantum computers~\cite{kjaergaard2020superconducting} using IBM Q Qiskit~\cite{Qiskit} or Google Cirq~\cite{cirq}.  To show the compatibility with qubits, we first show how to map our scheme onto spin-$1/2$ particles.  Second, we show that the key component of MITE, the measurement operators as given in Eq.~\eqref{eq:weakmeasurementoperator}, can be implemented using a circuit recompilation of the corresponding unitary operator.

We perform the mapping onto qubits via the transformation of the spin-$1$ operators as
  \begin{align}
    \label{eq:spinonetospinhalf}
    &\mathbf{S}_j \rightarrow \bm{\sigma}_{j,a}+\bm{\sigma}_{j,b}
  \end{align}
where the Pauli operators $ \bm{\sigma}_{j,l} $ is defined in Eq.~\eqref{spinhalf}.  The terms in Eq.~\eqref{eq:spin2projector} can be written in terms of Pauli matrices as
  \begin{align}
    \label{eq:P2inspinhalf}
    &\mathbf{S}_j \cdot \mathbf{S}_{j+1}\rightarrow \left(\bm{\sigma}_{j,a}+\bm{\sigma}_{j,b}\right) \cdot \left(\bm{\sigma}_{j+1,a}+\bm{\sigma}_{j+1,b}\right)
  \end{align}
Since each physical spin-$1$ site has been transformed to two spin-$1/2$ sites, the Hilbert space of the total chain has also been enlarged from $3^N$ to $2^{2N}=4^N$. By the usual rules of angular momentum coupling, the two spin-$1/2$ can couple to form a spin-$0$ and spin-$1$.  The states that are symmetric under interchange of the $a$ and $b$ spins form a spin-$1$ triplet, while the states that are antisymmetric form a spin-$0$ singlet.  Since the Hamiltonian is symmetric under $ a \leftrightarrow b $ interchange, and we will start in an initial state that is also symmetric
\begin{align}
  \label{eq:spinhalfinitialstate}
  &|\psi_0\rangle=\bigotimes_{j=1}^{N} |0 \rangle_{j,a} |0 \rangle_{j,b}   ,
\end{align}
the evolution remains in the symmetric subspace of dimension $ 3^N$, and the antisymmetric sector remains unpopulated throughout the subsequent evolution.

\begin{figure}[t]
	\centering
	\includegraphics[width=1.0\columnwidth,draft=false]{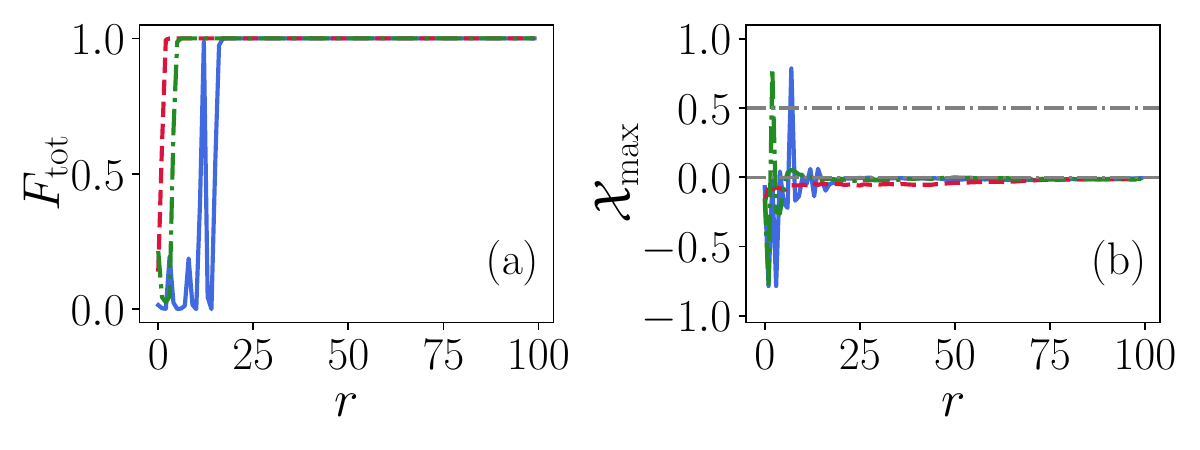}
	\caption{The AKLT state preparation in the spin-$1/2$ representation for $N=3$. (a) the total state fidelity $F_{\rm tot}$ as a function of MITE subroutine round $r$. For each $r$, MITE routines are performed throughout the whole system. (b) the peak value of the amplitude function as a function of time. We perform $100$ runs, and randomly choose three of them for the illustration in this figure, each color representing a single run. The data points are recorded within the iteration for each MITE subroutine round $r$ performed throughout the whole system up to $r=100$. For all panels, we have used $N_{\rm iter}=30$, and the interaction time is chosen to be $\epsilon=0.5$. The factor $\eta$ is chosen to be $\eta=2$ such that $E_{\text{th}} = 1/4 $.
  }
	\label{fig:N3statepreparation}
\end{figure}

Our numerical results for the MITE preparation of the AKLT state using in the qubit framework are shown in Fig.~\ref{fig:N3statepreparation}. 
Similarly to the spin-$1$ case, a high fidelity AKLT state can be obtained typically after $r \sim 10$ rounds of MITE subroutines performed throughout the whole system. Interestingly, the state appears to converge with a smaller number of rounds and the peak value $\chi_{\rm max}$ of the Gaussian function reaches the desired value faster than the spin-$1$ case, despite the mathematical equivalence of the models.  We note that although the Hilbert space for the spin-$1/2$ representation is larger than the case with spin-$1$ representation, 
due to the symmetry obeyed by the Hamiltonian and initial state, the evolution strictly remains in the symmetric sector and the additional Hilbert space available never affects the convergence. 

As with the $ S = 1$ case, for smaller $ \epsilon $ we find that the convergence is slower, and the peak value $\chi_{\rm max}$ fluctuates significantly before converges to the desired values, respectively. 

In summary, our MITE approach for generating the AKLT state stands out from MPS-based methods~\cite{Barratt2021parallel,Slattery2021quantum,Haghshenas2022variational,Miao2023quantum,malz2023preparation,Smith2022} by being a deterministic approach for spin-$1/2$, which has the potential to be implemented on a NISQ-era universal quantum computer. We want to highlight that it not only addresses the challenges associated with MPS approaches but also achieves a constant preparation time for the AKLT state, regardless of the chain length $N$.  Additionally, we remark that after the successful preparation of the AKLT state, subsequent procedures can be performed to directly measure observables, and calculate important quantities such as entanglement entropy~\cite{Choo2018EE} on a universal quantum computer.

\section{Implementation of the measurement operator}
\label{sec:impl}

The most straightforward implementation of the measurement operator [see Eq.~\eqref{eq:weakmeasurementoperator} from Methods] involves applying a Hamiltonian of the form 
\begin{align}
\hat{H}_M = \hat{\mathcal{P}}^{S=2}_{j,j+1} \otimes \sigma_a^{y} .
\end{align}
In the qubit language, the projector involves the square of the operator given in Eq.~\eqref{eq:P2inspinhalf}, which involve fourth order spin interactions.  Together with the ancilla qubit, this is a fifth order qubit interaction, which is likely to be difficult to implement using operations available on NISQ devices.  However, using the variational optimization algorithm for the quantum circuit recompilation~\cite{koh2022stabilizing,gray2018quimb,heya2018variational,khatri2019quantum,TanSun2021,cerezo2021variational,bittel2021training,KohLeeTopo2022,koh2023observation}, the measurement operator [Eq.~\eqref{eq:weakmeasurementoperator} from Methods] can be converted to a more readily implementable form.

\begin{figure}[t]
	\centering
	\includegraphics[width=1.0\columnwidth,draft=false]{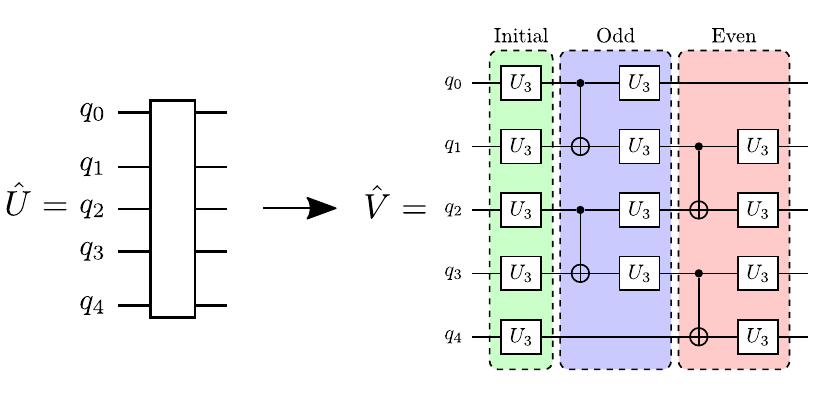}
	\caption{Recompiled quantum circuit for the measurement operator in Eq.~\eqref{eq:weakmeasurementoperator}. (left panel) The target unitary evolution operator $\hat{U}$; (right panel) the ansatz  $\hat{V}$ and its quantum circuit: an initial layer composed of $U_3$ gates followed by an odd layer and an even layer. The odd (even) layer has $CX$ gates on odd (even) bonds followed by $U_3$ gates placed on the corresponding physical sites. The operator $\hat{U}$ corresponds to two single spin-$1$ sites, where each site is transformed to two spin-$1/2$ sites as illustrated in measurement operator from Eq.~\eqref{eq:weakmeasurementoperator}, plus one additional ancilla qubit, thus rendering a total of $2 \times 2 + 1=5$-qubit unitary operator.}
	\label{fig:circuitrecompilation}
\end{figure}

In Fig.~\ref{fig:circuitrecompilation}, we substitute the original unitary operation for the measurement $\hat{U}= \exp( -i \epsilon \hat{H}_M ) $ into a variational parameterized circuit $\hat{V}$. This involves 5 qubits since the measurement is on two sites, which consist of two qubits each, and there is one ancilla qubit.  The parameterized circuit consists of an initial layer of $U_3$ gates, followed by alternate ``odd'' and ``even'' layers. Here the $U_3$ gate is defined in the same way as in Qiskit~\cite{Qiskit}: $U_3=U_3(\theta,\phi,\lambda)$. Each layer is composed of one layer of CNOT gates followed by a single layer of $U_3$ gates. The total depth of the parameterized circuit is $n_L$.  We use the \texttt{Quimb} package~\cite{gray2018quimb} to optimize the parametrized circuit $\hat{V}$ through a limited memory Broyden-Fletcher-Goldfarb-Shanno algorithm with box constraints (L-BFGSB)~\cite{koh2022stabilizing,Malouf2002,Cao2007}, with a basin-hopping method~\cite{li1987monte,Wales1997,Scheraga1999,wales_2004}. The loss function is the same as that given in Ref.~\cite{gray2018quimb}.   

\begin{figure}[h]
	\centering
	\includegraphics[width=1.0\columnwidth,draft=false]{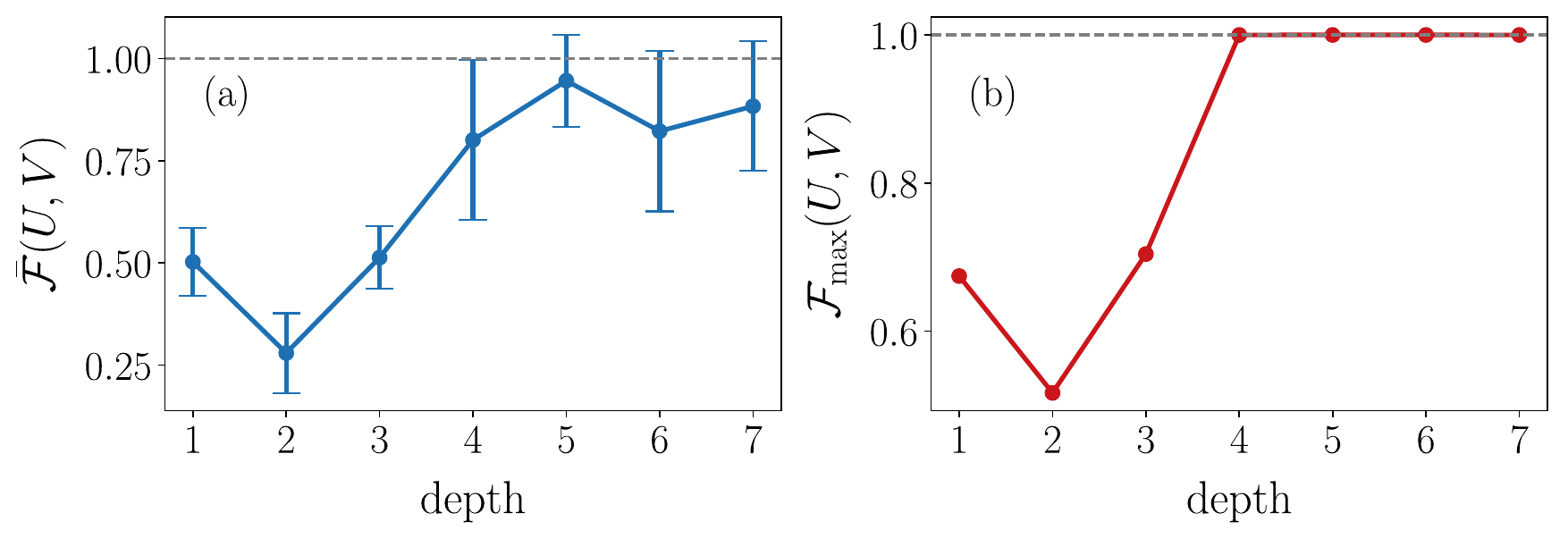}
	\caption{Circuit recompilation of the  measurement operator from Eq.~\eqref{eq:weakmeasurementoperator} for $N_M=2$. (a) the averaged fidelity between target operator $U$ and the ansatz circuit $V$. The average value and the error bar are calculated over $100$ repetitions. (b) the maximum of the fidelity between target operator $U$ and the ansatz circuit $V$ within every $100$ repetitions. Other parameters are: $N_{\rm iter}=100$, and $N_{\rm hop}=5$ from a basin-hopping method.}
	\label{fig:operatorcircuitrecompilation}
\end{figure}

The optimization results are shown in Fig.~\ref{fig:operatorcircuitrecompilation}, where we show both the averaged fidelity $\bar{\mathcal{F}}(U,V)$ as well as the maximum fidelity $\mathcal{F}_{\rm max}(U,V)$ between the original operator $\hat{U}$ and the parametrized circuit operator $\hat{V}$ of $100$ repetitions to characterize the performance of the optimization procedure. It is found that when the circuit layer depth is larger than $4$, of all the $100$ repetitions, the maximum fidelity between $\hat{U}$ and $\hat{V}$ already reaches the value larger than $99.99\%$ [Fig.~\ref{fig:operatorcircuitrecompilation}(a)]. When the depth is larger than $6$, although $\mathcal{F}_{\rm max}(U,V)$ remains above $99.99\%$, the average fidelity $\bar{\mathcal{F}}(U,V)$ starts to decrease due to over-parametrization and other basin-hopping parameters chosen. 

Hence in summary, we are able to approximate the unitary $\hat{U} $ with only a single-digit number of layers of $U_3$ and CNOT gates, where the number of CNOT gates is of the magnitude of $\sim 10$, and the fidelity between the original target weak measurement operator and the ansatz circuit is larger than $99.99\%$.  Using the default \texttt{transpile} function from Qiskit~\cite{Qiskit} for the same measurement operator, for this weak measurement operator corresponding to two spin-$1$ sites, the number of CNOT gates is 499.

\section{Discussion}
\label{sec:conclustions}

We have proposed an efficient quantum algorithm based on the measurement-based imaginary time evolution (MITE) algorithm for preparation of the AKLT state.  
Compared to MPS-based methods~\cite{Smith2022,Murta2022}, the MITE approach for preparing the AKLT state is deterministic, requiring no postselection. It addresses the limitations of MPS-based methods and reduces the preparation time to a constant with respect to the chain length $N$, despite the complexity of the quantum many-body wavefunction. Additionally, the weak measurement operator can be directly implemented on qubit-based devices.
Our results show that on average, there is a fast convergence to the target state typically with approximately $ T < 100$  
time steps per MITE subroutine. The total number of rounds is typically within $100$ for all system sizes considered. For the case of spin-$1/2$ representation, by using the variational optimization quantum circuit recompilation algorithm, the five-qubit weak measurement operator can be transformed into a circuit with $ \sim 10$ CNOT gates, which allows for a direct implementation on qubit-based quantum computers.  The speed of the overall convergence benefits  from a optimized interaction time parameter $ \epsilon $, if this is too small the convergence tends to occur much more slowly.

For a completely generic Hamiltonian, the complexity of convergence of the MITE algorithm typically scales with the dimension of the Hilbert space~\cite{MaoByrnes2022}.  However in the case of the AKLT state, due to the key property given in Eq.~\eqref{commonproj}, this allowed us to parallelize all MITE evolutions corresponding to even and odd $ j $ projectors $ \hat{\mathcal{P}}^{S=2}_{j,j+1} $.  Apart from the $ O(N) $ improvement in scaling, more importantly, since each projector only involves a convergence on a Hilbert space dimension of 2 sites (Hilbert space dimension of $ 3^2 = 9 $), the convergence of each MITE subroutine is independent of the chain length.  Remarkably, finding the the common eigenstate for all $ j $  in Eq.~\eqref{commonproj} is independent of the chain length, due to the short-ranged correlations in the AKLT state.  This was directly observed in our MITE simulations, as well as the prototypical case examined in Fig.~\ref{fig:convergence}.  This means that the whole convergence is attained in approximately constant time, which is an exponential speedup compared to the naive estimate of convergence related to the full Hilbert space dimension $ 3^N$.

We note that the efficiency of preparation observed here is a special case thanks to the properties of the AKLT state.  There are many other interesting states that could be also prepared using the  MITE approach, including exotic dissipative many-body physics~\cite{Ciuti2013,Sieberer2013,Joshi2013,Landa2020,Chen2020,Pistorius2020,Deuar2021,Chen2021,Naji2022,Jian2022,shen2023proposal}, as well as the simulation of quantum spin Hall effects beyond the solid-state 
systems~\cite{ByrnesDowling2015,ChenByrnes2019}.
We remark that since we utilized the properties of the AKLT state for the MITE algorithm, the question of whether there are other models that can be prepared with the same remarkable efficiency as the AKLT state remains unresolved.  Exploring this issue could lead to important insights into the underlying principles of quantum state preparation, particularly concerning the role of the Hamiltonian's structure in facilitating this exceptional efficiency. Thus, a deeper investigation is needed to determine if similar models exist and what exact properties they must possess to replicate or even surpass the efficiency observed in the AKLT state. It was shown in Ref.~\cite{KondappanByrnes2022} that cluster states could also be prepared efficiently due to the commuting stabilizers that define the state.  Even in the generic case, the MITE approach does not require any postselection, and is therefore deterministic: one could obtain the target AKLT state for every run of the sequence. Furthermore, no knowledge of the ground state wavefunction is required for the convergence.  Hence it is potentially useful as a generic procedure in a quantum simulation context, where eigenstates of Hamiltonians require preparation.

\section*{Acknowledgments}
T.~B. is supported by the National Natural Science Foundation of China (62071301); NYU-ECNU Institute of Physics at NYU Shanghai; the Joint Physics Research Institute Challenge Grant; the Science and Technology Commission of Shanghai Municipality (19XD1423000,22ZR1444600); the NYU Shanghai Boost Fund; the China Foreign Experts Program (G2021013002L); the NYU Shanghai Major-Grants Seed Fund; Tamkeen under the NYU Abu Dhabi Research Institute grant CG008; and the SMEC Scientific Research Innovation Project (2023ZKZD55). T.~C. acknowledges support from the Singapore National Research Foundation (NRF) via the Project No.~NRF2021-QEP2-02-P09, and No.~NRF-NRFF12-2020-0005, as well
as the support by the NRF, Singapore and A*STAR under its CQT Bridging Grant. The computational work for this article was supported in part through the NYU IT High Performance Computing resources, services, and staff expertise, and was partially performed on resources of the National Supercomputing Centre, Singapore (NSCC) (\url{https://www.nscc.sg}), and on resources of the National University of Singapore (NUS)'s high-performance computing facilities, partially supported by NUS IT's Research Computing group.

\bibliographystyle{quantum}


\providecommand{\noopsort}[1]{}\providecommand{\singleletter}[1]{#1}%


\end{document}